\documentclass{iopart}
\usepackage{iopams, setstack}
\usepackage{bm,hyperref}

\newcommand{\eqref}[1]{{(\ref{#1})}}    

\newcommand{\ItP}{{\mathcal{P}}} 

\newcommand{\Lie}{{\mathcal{L}}} 
\newcommand{\LieX}{{\mathcal{L}_{\xi}}} 

\begin{document}
\title{Electromagnetic self-forces and generalized Killing fields}
\author{Abraham I. Harte}
\address{Enrico Fermi Institute}
\address{University of Chicago, Chicago, IL 60637 USA}
\ead{harte@uchicago.edu}

\begin{abstract}
Building upon previous results in scalar field theory, a formalism is developed that uses generalized Killing fields to understand the behavior of extended charges interacting with their own electromagnetic fields. New notions of effective linear and angular momenta are identified, and their evolution equations are derived exactly in arbitrary (but fixed) curved spacetimes. A slightly modified form of the Detweiler-Whiting axiom that a charge's motion should only be influenced by the so-called ``regular'' component of its self-field is shown to follow very easily. It is exact in some interesting cases, and approximate in most others. Explicit equations describing the center-of-mass motion, spin angular momentum, and changes in mass of a small charge are also derived in a particular limit. The chosen approximations -- although standard -- incorporate dipole and spin forces that do not appear in the traditional Abraham-Lorentz-Dirac or Dewitt-Brehme equations. They have, however, been previously identified in the test body limit.
\end{abstract}

\vskip 2pc

\section{Introduction}

Interactions between compact classical charge distributions and their own electromagnetic fields have now been studied in various contexts for more than a century \cite{Abraham, Schott, Dirac, DewittBrehme, Crowley, Havas, KijowskiSelfForce, QuinnWald, DetWhiting, PoissonRev, OriExtend, HarteEM, EFTSelfForce, ChargedBH, BobSamMe}. The underlying physics is well-understood. Maxwell's equations are to be coupled to some version of continuum mechanics describing the type of matter under consideration. The resulting partial differential equations are then used to evolve the system forward in time from some appropriate set of initial data. Unfortunately, this proposal almost always requires the use of numerical methods. These are both difficult to implement and obscure general trends that may exist across classes of solutions.

A trivial example of such a trend is found in the motion of sufficiently small charge distributions in the test body limit. To an excellent degree of precision, the center-of-mass position of such a body is governed by a single ordinary differential equation depending on only one parameter -- the (conserved) charge-to-mass ratio. No other details of the object's composition need to be taken into account in order to characterize its bulk motion. This is not to say that they are completely irrelevant. Even in this highly restricted regime, the \textit{internal} dynamics are not in any sense universal. Regardless, focusing attention on special variables such as the center-of-mass position can lead to interesting and relatively simple analytic results.

The example just discussed may be extended in several ways. Consider relaxing the constraints on a body's physical size (with respect to the wavelengths of externally-imposed fields) while remaining in the test body regime. Doing so, one finds excellent approximations for the motion -- coupled to equations also describing the evolution of the mass and spin angular momentum -- as ordinary differential equations involving a small number of parameters. These are the multipole moments of the charge distribution. The full theory of such effects has been developed by Dixon \cite{Dix67, Dix70a, Dix74, Dix79}. It holds even in strongly curved spacetimes, and is adequate for describing almost all isolated charges of practical interest.

Still, there are some cases where a body's local field must be taken into account. Self-force or radiation reaction effects usually become significant only when changes in a charge's structure or overall motion occur on timescales approaching its light-crossing time. This might be expected in the highly-magnetized regions surrounding neutron stars and other extreme phenomena. Also of astrophysical interest is the gravitational analog of this problem. Consider, for example, a stellar-sized compact mass spiralling into a supermassive black hole. The gravitational radiation emitted in this process is expected to be observationally significant in the near future, so an improved theoretical understanding is vital. Considerable success has already been achieved in this area \cite{PoissonRev, MST, BobSam}, although some aspects might be better understood by first considering improved electromagnetic models.

Regardless of motivation, this paper considers the problem of a compact charge distribution interacting with its own electromagnetic field (as well as those of distant sources). Despite the extensive literature on the subject, there are lingering questions. One of these regards the surprising success of point particle methods. It is commonly known that there are no solutions to the equations coupling Maxwell's equations to sources with vanishing radius but finite mass and charge. Despite this, various methods have been proposed to remove problematic infinities and obtain equations of motion that agree with those derived by considering well-defined limits of extended charge distributions \cite{KijowskiSelfForce, QuinnWald, DetWhiting}. While there are intuitive arguments supporting these prescriptions, their connection to the full theory (involving extended charges) is unclear. The precise assumptions required to recover a physically and mathematically sensible ``point particle limit'' are not obvious, and should be better understood.

The fact that point particle methods work at all is suggestive, however.  Direct treatments of the dynamics of extended bodies have usually involved extremely tedious calculations in perturbation theory \cite{Schott, Crowley, OriExtend, HarteEM}. This has been especially true when allowing for arbitrary shapes and internal dynamics. The calculations used to obtain equations of motion for point particles are considerably simpler. This suggests that at least some of their hypotheses might have analogs in the theory of extended charges.

We demonstrate this for the so-called Detweiler-Whiting axiom. Drawing on ideas originally developed by Dirac in flat spacetime \cite{Dirac}, it states that a point charge only moves in response to a particular component of its self-field satisfying the vacuum Maxwell equations \cite{DetWhiting}. Applying this hypothesis very quickly leads to the Dewitt-Brehme (or generalized Abraham-Lorentz-Dirac) equation \cite{DewittBrehme} typically expected to describe the motion of small charges in a fixed background spacetime. We show that a simple non-perturbative calculation allows an analogous statement to be derived from the full continuum theory. Furthermore, the ``ignored'' portion of a charge's self-field is shown to act mainly as an effective shift in its linear and angular momenta.

Although self-force problems are typically thought of as intrinsically perturbative, several results derived here are exact up to effects related to the charge's gravitational self-field. They are not immediately sufficient to compute the center-of-mass behavior, however. The main difficulty lies in obtaining suitable explicit solutions to Maxwell's equations. Effects of the spacetime curvature can also be difficult to take into account for large bodies. An approximation procedure is therefore applied in order to obtain explicit equations of motion. It shrinks the charge's size while maintaining its charge-to-mass ratio and relative self-energy. The shape and internal structure are otherwise arbitrary except for the restriction that internal timescales do not shorten as the radius is decreased. Under these assumptions, equations describing the evolution of the body's mass, angular momentum, and center-of-mass position are derived. They effectively add Dewitt-Brehme-type self-forces to effects already expected for a test body up to dipole order.

The methods used here generalize those recently developed for understanding extended bodies coupled to Klein-Gordon fields \cite{HarteScalar}. Roughly speaking, linear and angular momenta are first defined in curved spacetime by introducing a 10-parameter family of vector fields that generalize the Killing symmetries of Minkowski spacetime \cite{HarteSyms}. These ideas are reviewed in section \ref{Sect:Prelims}. Various contributions to the evolution of the body's momenta are then analyzed in section \ref{Sect:SelfInt}. A particular component of the self-field is shown to act almost entirely as an effective inertia. Its remaining effect involves changes in an underlying Green function under flows defined by the generalized Killing fields. This has an intuitive interpretation as a violation of Newton's third law. It is shown that there are many interesting cases where such violations remain negligible or even vanish entirely. Finally, section \ref{Sect:EqsMotion} derives equations of motion for a small slowly-varying charge in curved spacetime.

\section{Preliminaries}
\label{Sect:Prelims}

If the only long-range dynamical field in a system is
electromagnetic, matter distributions that couple to it might be modeled by an action
\begin{equation}
  S = S_{\mathrm{M}} + \int (J^a A_a -\frac{1}{16 \pi} F^{ab} F_{ab}) \rmd V
  .
  \label{ActionMaxwell}
\end{equation}
The field strength $F_{ab} = 2 \nabla_{[a} A_{b]}$ is to be derived from the potential $A_a$ in the usual way, while the matter action $S_{\mathrm{M}}$ is assumed to be independent of these variables. Subtle issues regarding the proper formulation of electrodynamics in permeable and polarizable media will not be considered in any detail here \cite{DeGroot, AbrahamMinkRev, PenfieldHaus, IsraelPol}. It will be assumed that all such effects can be modeled by decomposing the current $J^a$ into appropriate ``microscopic'' and ``macroscopic'' components. Taking \eqref{ActionMaxwell} as a given, the standard field equation
\begin{equation}
  \nabla^b F_{ab} = 4 \pi J_a
  \label{MaxwellEq}
\end{equation}
follows immediately after varying the action with respect to the potential.

Diffeomorphism invariance further implies that the system's total stress-energy tensor is conserved. It is useful to define a stress-energy tensor $T_{ab}$ only for the charged body under consideration. Let
\begin{equation}
  T_{ab} = - \frac{2}{\sqrt{-g}} \frac{\delta S}{ \delta g^{ab} } -  T_{ab}^{\mathrm{EM}}  , \label{TDef}
\end{equation}
where
\begin{equation}
  T_{ab}^{\mathrm{EM}} = \frac{1}{4\pi} ( F_{a}{}^{c} F_{bc} -\frac{1}{4} g_{ab} F^{cd} F_{cd} ) \label{TEM}
\end{equation}
is the field's stress-energy tensor. $\delta S/ \delta g^{ab}$ represents a functional derivative of the action with respect to the inverse metric $g^{ab}$. The support of $T_{ab}$ is assumed to be a timelike worldtube $W$ that may be foliated with a set of compact spacelike hypersurfaces. Total stress-energy conservation now requires that the body's energy, momenta, and stress satisfy
\begin{equation}
  \nabla_b T^{ab} = F^{ab} J_b .
  \label{StressCons}
\end{equation}
The right-hand side of this equation may be interpreted as the generic force density exerted by an electromagnetic field on matter with current density $J^a$.

In general, much more can be learned from \eqref{ActionMaxwell}.
$S_{\mathrm{M}}$ implicitly contains a number of dynamical variables
describing the detailed state of the material under consideration. Their behaviors can be
extracted by applying the least action principle to each in turn. It
is then found that systems described by the given action obey a
coupled set of (usually) nonlinear partial differential equations.
There are very few interesting cases where a direct analytical
solution is feasible to obtain. Furthermore, these equations and their
solutions both depend strongly on the type of material under
consideration -- i.e., the detailed form of $S_{\mathrm{M}}$. Put another way, the full equations of motion always imply stress-energy conservation. The converse is almost never true. Despite this, it is known that there are cases where the details of a material's internal structure do not strongly affect its motion as a whole. Certain degrees of freedom can be almost entirely constrained by universal conservation laws like \eqref{StressCons}.

Such laws are related to the underlying geometry (and the internal gauge symmetry, which only implies charge conservation here). As is well-known, the continuous isometries of Minkowski spacetime imply natural definitions for
conserved quantities typically referred to as the body's linear and
angular momenta. Somewhat unconventionally, we choose to view these objects as maps from the space of Killing fields into $\mathbb{R}$. This is useful in that both the linear and angular momenta may be considered simultaneously. All quantities of interest are also scalars rather than first or second rank tensor fields.

This scheme may be extended to curved spacetimes that do not admit any Killing vectors. Given some reference worldline $\Gamma$ and an associated foliation $\Sigma(s)$, it is always possible to introduce
something that may be interpreted as a generalized Poincar\'{e} group $GP$\footnote{This is only possible in a neighborhood of $\Gamma$. It requires that any point in $W \cap \Sigma(s)$ be connected to $\Gamma \cap \Sigma(s)$ by a single (spacelike) geodesic. This essentially means that each spatial section of the body's worldtube is assumed to be within a normal neighborhood of the ``origin.'' Spacelike geodesics are not easily focused, so this is a very minor restriction.} \cite{HarteSyms}. All of its elements may be represented by linear combinations of $10$ independent vector fields in the four spacetime dimensions assumed here. By analogy to the situation in Minkowski (or de Sitter) spacetime, denote a body's linear and angular momenta by a map $\ItP_\xi(s) :
GP \times \mathbb{R} \rightarrow \mathbb{R}$. Let
\begin{equation}
  \ItP_\xi(s) = \int_{\Sigma(s)} T^{a}{}_{b} \xi^b \rmd S_a,
    \label{PDef}
\end{equation}
where $\xi^a \in GP$ is a generalized Killing field (GKF)
\cite{Dix70a, Dix74, HarteSyms}. Other reasonable definitions for $\ItP_\xi$ exist. Terms may be added to the right-hand side of \eqref{PDef} involving $J^a$ and $F_{ab}$, for example \cite{Dix70a, Dix74}. This possibility is discussed in the appendix. Furthermore, an additional effective momentum due to the body's self-field will be derived below. $\ItP_\xi$ should therefore be thought of more as an initial guess than a fundamental definition.

Relatively few properties of the GKFs are needed here. We note only
that any element of $GP$ is uniquely fixed by specifying $\xi^a$
and $\nabla_a \xi_b = \nabla_{[a} \xi_{b]}$ at any point $\gamma \in
\Gamma$ on the reference worldline. The full generalized Killing fields depend linearly (and non-degenerately) on these data. They also satisfy
\begin{equation}
  \LieX g_{ab} |_\Gamma = \nabla_a \LieX g_{bc} |_\Gamma = 0 .
  \label{Lieg0}
\end{equation}
Any exact Killing fields that may exist are necessarily GKFs. A precise definition for these objects and an extensive discussion of their properties may be found in
\cite{HarteSyms}.

The map $\ItP_\xi$ is easily related to more standard definitions
of linear and angular momenta. Let these objects be tensor fields
$p_a(s)$ and $S^{ab} = S^{[ab]}(s)$ on $\Gamma$. Then suppose that
\begin{equation}
  \ItP_\xi(s) = (p_a \xi^a + \frac{1}{2} S^{ab} \nabla_{[a} \xi_{b]})_{\gamma(s)} \label{pToP}
  ,
\end{equation}
where the point $\gamma(s) = \Gamma \cap \Sigma(s)$. The momenta are uniquely defined by this equation if $\ItP_\xi$ is known for all possible
GKFs. They are exactly the expected tensor fields in flat spacetime. Similar equations are also useful in Newtonian physics \cite{HarteScalar}.

\section{Self-interaction}
\label{Sect:SelfInt}

The motion of a compact charge is influenced by external
electromagnetic fields as well as its own. While it is very
difficult to compute every detail of the resulting dynamics, the
momenta are relatively simple to evolve. Consider changes in
$\ItP_\xi(s)$ between two times $s = s_1$ and $s = s_2 > s_1$.
Using \eqref{StressCons} and \eqref{PDef},
\begin{equation}
    \fl  \qquad \quad ( \ItP_\xi ) |_{s_1}^{s_2} = \ItP_\xi(s_2) - \ItP_\xi(s_1) =
    \int_{\Omega(s_1,s_2)} ( \xi^a J^b F_{ab} + \frac{1}{2} T^{ab} \LieX
    g_{ab}) \rmd V . \label{TotForce}
\end{equation}
Here, $\Omega(s_1,s_2)$ is defined to be that portion of the worldtube bounded by
the hypersurfaces $\Sigma(s_1)$ and $\Sigma(s_2)$. It is trivial to derive from this an expression for the instantaneous momentum change $\dot{\ItP}_\xi = \rmd \ItP_\xi / \rmd s$, although the averaged form given here will be more useful for now.

The term on the right-hand side of \eqref{TotForce} involving $\LieX
g_{ab}$ represents the covariant gravitational force and torque on
a compact mass. It exists even in the absence of any electromagnetic fields. Assuming that the body does not significantly influence the metric, $g_{ab}$ can usually be assumed to vary slowly inside the worldtube in (say) an appropriate normal coordinate system. It is then possible to perform a Taylor expansion and evaluate the relevant integrals term by term. Given \eqref{Lieg0}, it is clear that the lowest-order approximation for the gravitational momentum shift involves the quadrupole moment of $T^{ab}$. Full multipole expansions have been given in \cite{Dix74, Dix79}. Specific examples of the dynamics arising from taking into account only the dominant quadrupole term have recently been discussed in several contexts \cite{HarteQuadrupole, ItalianQuadrupole, ItalianQuadrupole2}.

The focus here will be on the electromagnetic term appearing in
\eqref{TotForce}. This may first be simplified by splitting up the field
into portions sourced by the charge itself and by the external
universe:
\begin{equation}
  F_{ab} = F_{ab}^{\mathrm{ext}} + F_{ab}^{\mathrm{self}} .
\end{equation}
The self-field must satisfy
\begin{equation}
  \nabla^b F_{ab}^{\mathrm{self}} = 4 \pi J_a . \label{MaxwellEqSelf}
\end{equation}
The current density here is assumed to vanish outside of $W$. All contributions from distant charges are therefore ignored in this equation. They are included in \eqref{MaxwellEq}, however. Suppose that the self-field is defined by solving \eqref{MaxwellEqSelf} with retarded boundary conditions. It follows that
\begin{equation}
  F_{ab}^{\mathrm{self}} = 2 \int_W \nabla_{[a} G_{b]b'}^{\mathrm{ret}} J^{b'} \rmd
  V' , \label{SelfField}
\end{equation}
where $G_{aa'}^{\mathrm{ret}}(x,x')$ is the retarded Green function
for the vector potential in some (as-yet unspecified) gauge. These
definitions imply that the external field is a solution to the
homogeneous Maxwell equations at least in a neighborhood of $W$. It will often vary slowly throughout the body, so the forces and torques that it exerts may be easily approximated using series that involve successively higher multipole
moments of $J^a$.

The influence of $F_{ab}^{\mathrm{self}}$ cannot be evaluated so
directly. It will vary on scales at least as small as the body's
diameter, so multipole expansions are not immediately helpful.
Despite this, certain limits are expected to exist where the
self-force becomes proportional to the square of the total charge
$q$. This simple result -- involving only the monopole moment of the
charge distribution -- suggests that there is a sense in which a
simple multipole expansion is indeed applicable. The absence of
self-forces and self-torques in Newtonian electrostatics (without dielectrics) leads one to
expect that only a small portion of the field sourced by a
relativistic body will directly affect its momenta. It is then
reasonable to hypothesize that the remainder might vary so slowly
as to exert forces and torques depending mainly on the body's
first few multipole moments. This idea has been formalized by the
Detweiler-Whiting axiom \cite{DetWhiting}, which states that the
motion of a point particle is only affected by a particular
component of its self-field satisfying the homogeneous Maxwell
equations.

One of the main goals of this work is to write the laws of motion
for extended bodies in a form where such results are essentially
self-evident in an appropriate limit. The `negligible' portion of
the self-field is identified and (largely) removed at the outset,
leaving only a nearly-homogeneous piece that produces the dominant
self-force. This program first requires splitting the self-field
into two components:
\begin{equation}
F_{ab}^{\mathrm{self}} = F_{ab}^{\mathrm{S}} + F_{ab}^{\mathrm{R}}.
\label{FRSDef}
\end{equation}
By convention, the quantities on the right-hand side are
respectively referred to as the singular and regular components of
the physical self-field. These names refer to the respective
behaviors of the two fields when associated with a point particle.
Charge distributions considered here are assumed to be smooth,
so all of their associated fields remain bounded. $F_{ab}^{\mathrm{S}}$ is
intended to include most of the detailed structure of the self-field
while having little effect on the body's overall motion. It will
contribute an effective inertia, but does not otherwise influence
the momenta as strongly as $F_{ab}^{\mathrm{R}}$. One might expect
this component of the self-field to be approximately Coulombian in
form, and to be `bound' to the particle in some sense.

The regular component of the self-field is to be constructed so as
to vary slowly inside the body under reasonable conditions. We
therefore suppose that it satisfies the homogeneous Maxwell
equations. Let
\begin{equation}
  \nabla^b F_{ab}^{\mathrm{R}} = 0; \qquad \nabla^b F_{ab}^{\mathrm{S}} = 4 \pi J_a . \label{MaxwellRS}
\end{equation}
Also demand that, for example, $F_{ab}^{\mathrm{S}} = 2 \nabla_{[a} A_{b]}^{\mathrm{S}}$ for an appropriate vector potential $A_{a}^{\mathrm{S}}$. Suppose that all such potentials are to be derived from Green functions in the simplest possible manner; e.g.
\begin{equation}
  A_a^{\mathrm{R}} = \int_W G_{aa'}^{\mathrm{R}} J^{a'} \rmd V' . \label{ADef}
\end{equation}
Unless otherwise noted, we adopt the Lorenz gauge throughout this paper. The Green functions are then known to satisfy
\begin{equation}
  \nabla^b \nabla_b G_{aa'}^{(\cdots)} - R_{a}{}^{b} G^{(\cdots)}_{ba'} = - 4 \pi g_{aa'} \delta(x,x') , \label{GreenLorenz}
\end{equation}
where $g_{aa'}(x,x')$ is any two-point tensor field that reduces to the metric as $x \rightarrow x'$. For concreteness, let it be the parallel propagator. Note that vector potentials obtained by substituting a Lorenz-gauge Green function into \eqref{ADef} are only solutions to Maxwell's equations when $\nabla^a A_a = 0$. Starting from the identity
\begin{equation}
  \nabla_a [ g^{a}{}_{a'} \delta(x,x')] = - \nabla_{a'} \delta(x,x') ,
\end{equation}
it may be shown that \cite{DewittBrehme}
\begin{equation}
  \nabla_{[a'} ( \nabla^a G^{(\cdots)}_{|a|b']} ) = 0 .
\end{equation}
The gauge condition is then seen to follow as long as $\nabla_a J^a = 0$.


\subsection{Detweiler-Whiting Green functions}

A specific form for the singular self-field is already expected from the Detweiler-Whiting axiom. The appropriate Green functions are defined in Lorenz gauge, so they satisfy equations like \eqref{GreenLorenz}. Several auxiliary conditions effectively fix the boundary conditions, and therefore the solutions. First, the regular and singular Green functions are to be related to the retarded Green function exactly as might be expected from \eqref{FRSDef}:
\begin{equation}
  G_{aa'}^{\mathrm{ret}}(x,x') = G_{aa'}^{\mathrm{R}}(x,x') + G_{aa'}^{\mathrm{S}}(x,x') .
\end{equation}
The left-hand side vanishes when the source point $x'$ is not in the causal past of the field point $x$. We restrict the S-type Green function from having support in the timelike past or future of any field point. This motivates its identification as the propagator for a bound field able to freely transfer momentum to or from the matter. Computing the singular field at a given point only requires knowledge of the charge distribution a short finite time into the past (and future).

The last condition placed on the singular Green function requires it to be symmetric in its arguments:
\begin{equation}
  G^{\mathrm{S}}_{aa'}(x,x') = G^{\mathrm{S}}_{a'a}(x,x') . \label{GSym}
\end{equation}
This is necessary in order for there to be a sense in which individual components of the singular self-force approximately obey Newton's third law (see section \ref{Sect:Newt3rd}). The given conditions uniquely determine $G_{aa'}^{\mathrm{R}}$ and $G_{aa'}^{\mathrm{S}}$, at least in some finite region. They are referred to as the Detweiler-Whiting Green functions. In flat spacetime, $G_{aa'}^{\mathrm{S}} = (G_{aa'}^{\mathrm{ret}} + G_{aa'}^{\mathrm{adv}})/2$ and $G_{aa'}^{\mathrm{R}} = (G_{aa'}^{\mathrm{ret}} - G_{aa'}^{\mathrm{adv}})/2$.

More generally, these distributions can be at least partially written down with the help of Synge's world function $\sigma(x,x')$.  This two-point scalar represents one-half of the squared geodesic distance between its arguments. The support and symmetry requirements placed on the S-type Green function guarantee that it has the form
\begin{equation}
  G^{\mathrm{S}}_{aa'} = \Delta_{aa'} \delta( \sigma ) + V_{aa'} \Theta ( \sigma ) ,
  \label{GExpand}
\end{equation}
where $\Delta_{aa'} = \Delta_{a'a}$ and $V_{aa'} = V_{a'a}$ are appropriate bitensors to be determined by substitution into \eqref{GreenLorenz}. $\Theta(\sigma)$ is a unit step function that vanishes when $\sigma < 0$ (i.e., when $x$ and $x'$ are timelike-separated). The singular portion of \eqref{GExpand} is not difficult to express entirely in terms of $\sigma$ and the parallel propagator $g^{a}{}_{a'}$. First introduce the (scalarized) van Vleck determinant
\begin{equation}
  \Delta(x,x') = - \frac{ \det[ - \nabla_a \nabla_{a'} \sigma(x,x') ] }{ \sqrt{-g} \sqrt{-g'} } .
\end{equation}
It may then be shown that \cite{PoissonRev}
\begin{equation}
  \Delta_{aa'} = \frac{1}{2} g_{aa'} \Delta^{1/2} ,
\end{equation}
where, again, $g_{aa'}$ is the parallel propagator. The remaining ``tail'' portion of the Green function cannot be evaluated so easily. A separate analysis must be carried out in each spacetime. $V_{aa'}$ very rarely vanishes, although there are interesting cases -- conformally-flat spacetimes, for example -- where its existence is purely a gauge effect \cite{Tails}. This contrasts sharply with scalar field theory, where physically-relevant tails almost always exist.

\subsection{Newton's third law}
\label{Sect:Newt3rd}

It has been emphasized that the singular self-field is to be constructed so as to have a minimal effect on the charge's motion. Taking cues from the non-relativistic theory, one might expect to arrange this by ensuring that forces generated by $F_{ab}^{\mathrm{S}}$ satisfy an approximate form of Newton's third law. This procedure is considerably more ambiguous than in the scalar case described in \cite{HarteScalar}, although it still provides useful motivation.

Consider the force or torque exerted by one portion of the body on another and vice versa. This requires several clarifications. First, forces and torques are typically represented as vectors and 2-forms, respectively. The force on (say) one element of charge therefore cannot be immediately compared with the force on another element. This is resolved in exactly the same way that the linear and angular momenta were absorbed into a set of scalars via \eqref{PDef} and \eqref{pToP}. A particular component of force or torque can be picked out by choosing an appropriate GKF. Differentiating \eqref{pToP} and using \eqref{Lieg0} shows that
\begin{equation}
  \rmd \ItP_\xi / \rmd s = \dot{\ItP}_\xi = F^a \xi_a + \frac{1}{2} N^{ab} \nabla_{[a} \xi_{b]} ,
\end{equation}
where the net force $F^a$ and torque $N_{ab} = N_{[ab]}$ are defined by
\begin{equation}
  F^a = \dot{p}^a - \frac{1}{2} S^{bc} \dot{\gamma}^d R_{bcd}{}^{a} ; \quad N_{ab} = \dot{S}_{ab} - 2 p_{[a} \dot{\gamma}_{b]}. \label{ForceDef}
\end{equation}
There is no loss of generality in considering $\dot{\ItP}_\xi$ for all possible GKFs instead of $F^a$ and $N_{ab}$ themselves. The generalized force acting on a particular component of the body may therefore be identified with its contribution to changes in the generalized momentum $\ItP_\xi$.

Let $\mathcal{F}_\xi (x,x')$ denote just such a generalized force density acting in a `direction' $\xi^a$ on charge in a small spacetime region $\rmd V$ due the singular field sourced by charge in $\rmd V'$. Given \eqref{TotForce}, it might be reasonable to define this by
\begin{equation}
  \mathcal{F}^{\mathrm{S}}_\xi(x,x') \rmd V' = \xi^a J^b
  F_{ab}^{\mathrm{S}} [\rmd V'] , \label{ForceDensity}
\end{equation}
where $F_{ab}^{\mathrm{S}}[\rmd V']$ provides some notion of the field arising from charge in $\rmd V'$. It is perhaps most intuitive to define the field sourced by charge in a spacetime volume $\Lambda$ by
\begin{equation}
  F_{ab}^{\mathrm{S}}[\Lambda] = 2 \int_\Lambda
  \nabla_{[a} G_{b]a'}^{\mathrm{S}} J^{b'} \rmd V' , \label{BracketNot}
\end{equation}
so $F_{ab}^{\mathrm{S}} [\rmd V'] = 2 \nabla_{[a} G^{\mathrm{S}}_{b]b'} J^{b'} \rmd V'$. Note that the effective source in \eqref{BracketNot}
is usually not conserved when $W \nsubseteq \Lambda$. The
associated field strength is therefore not a solution to Maxwell's
equations with any physically-relevant charge distribution. The field $F_{ab}[\rmd V']$ apparently associated with charge only at a single spacetime point is actually generated by an effective current density
\begin{equation}
  \hat{J}_a = J_a \delta(x,x') + \frac{1}{4\pi} \nabla_a ( \nabla^b G^{\mathrm{S}}_{bb'} ) J^{b'} .
\end{equation}
Despite these interpretational difficulties, the given constructions are still suggestive. Varying
over all possible GKFs  in \eqref{ForceDensity} allows ordinary
forces and torques to be recovered at least up to total
divergences. Including additional terms that take into account the inherent ambiguities does not significantly alter the discussion.

In terms of the force density $\mathcal{F}_\xi$, the influence of the S-type electromagnetic self-field in \eqref{TotForce} on the body's motion is easily shown to have the form
\begin{eqnarray}
  \int_\Omega \rmd V \xi^a J^b  F_{ab}^{\mathrm{S}}  =& \frac{1}{2}
  \int_\Omega \rmd V \bigg( \int_W \rmd V' \left[ \mathcal{F}_\xi^{\mathrm{S}} (x,x')+\mathcal{F}_\xi^{\mathrm{S}} (x',x) \right]
  \nonumber
  \\
  & \quad ~ + \int_{W \setminus \Omega} \rmd V' \left[ \mathcal{F}_{\xi}^{\mathrm{S}} (x,x') - \mathcal{F}_{\xi}^{\mathrm{S}} (x',x) \right]  \bigg) .
  \label{ForceExpand}
\end{eqnarray}
The arguments of $\Omega = \Omega(s_1,s_2)$ have been dropped here for simplicity. Note that the first pair of terms on the right-hand side essentially measure the failure of Newton's third law. Using \eqref{GSym} and \eqref{ForceDensity} shows that
\begin{eqnarray}
    \mathcal{F}_\xi^{\mathrm{S}}(x,x') +
    \mathcal{F}_\xi^{\mathrm{S}}(x',x) &=& J^b J^{b'} \LieX G_{bb'}^{\mathrm{S}} - \nabla_a (J^a \xi^b J^{b'}
    G_{bb'}^{\mathrm{S}})  \nonumber
    \\
    && ~ - \nabla_{a'} ( J^{a'} \xi^{b'} J^b
    G_{bb'}^{\mathrm{S}}) ,
    \label{Newt3rd}
\end{eqnarray}
where Lie derivatives of two-point tensors are defined to act individually on both arguments. If this result is averaged over finite volumes, the contributions from the two total divergences turn into boundary integrals that are easily dealt with. The first term more directly measures the degree to which action-reaction pairs are unbalanced. It is all that appears in the scalar field theory \cite{HarteScalar}.
Importantly, the properties of the generalized Killing fields
ensure that it remains negligible in many interesting regimes. It
is also clear that it vanishes entirely if $\xi^a$ is exactly
Killing. This is a very useful consequence of adopting Lorenz gauge. Other choices can lead to Green functions that are not invariant under any Killing fields that may exist.

\subsection{Laws of motion}

It is evident from this discussion that a significant simplification may occur if the only direct influence of the self-field can be shown to involve $\LieX G_{aa'}^{\mathrm{S}}$. This actually occurs quite naturally. Returning to \eqref{ForceExpand}, the left-hand side may be interpreted as the integrated force or torque acting on the body due to $F_{ab}^{\mathrm{S}}$ between times $s_1$ and $s_2$. It would therefore be reasonable to expect the integrand on the right-hand side to have essentially the same form on each time slice. More precisely, the integrand of the outer integral (over $\Omega$) in \eqref{ForceExpand} should not explicitly depend on $s_1$ or $s_2$. It does do so, however. This is a consequence of having used the elementary identity \begin{equation}
  \int_\Omega \rmd V \int_W \rmd V' \mathcal{A}(x,x') = \int_\Omega \rmd V \int_{W \setminus \Omega} \rmd V' \mathcal{A}(x,x'),
\end{equation}
which holds for any sufficiently well-behaved $\mathcal{A}(x,x') = - \mathcal{A}(x',x)$.

The $s$-dependent terms can be interpreted as effective inertias. To see this, temporarily suppose that $s_2 - s_1$ is sufficiently large that any element of $\Sigma(s_2)$ may be connected to any element of $\Sigma(s_1)$ by a timelike curve. Roughly speaking, this means that we're considering time differences larger than the body's diameter. The assumed support properties of $G_{aa'}^{\mathrm{S}}$ then imply that the second line of \eqref{ForceExpand} becomes a double integral over points in $W$ within a light-crossing time of $\Sigma(s_1)$ and $\Sigma(s_2)$. It does not depend in any way on the behavior of the body in the distant past or future, nor in the ``middle'' of $\Omega$. Furthermore, the contributions from regions near $\Sigma(s_1)$ and $\Sigma(s_2)$ can be separated into a difference of two integrals with identical integrands (they differ only in the region of integration). Combining \eqref{TotForce}, \eqref{ForceExpand}, and \eqref{Newt3rd},
\begin{eqnarray}
  \fl (\ItP_\xi + \mathcal{E}_\xi)|_{s_1}^{s_2} = \int_\Omega \rmd V \bigg[ \frac{1}{2} T^{ab} \LieX g_{ab} + \xi^a J^b  (F_{ab}^{\mathrm{ext}} + F_{ab}^{\mathrm{R}}) + \frac{1}{2} \int_W \rmd V' J^a J^{a'} \LieX G_{aa'}^{\mathrm{S}} \bigg], \label{FinalDiscrete}
\end{eqnarray}
where
\begin{equation}
    \fl \qquad \mathcal{E}_\xi(s) = \frac{1}{2} \bigg( \int_{\Sigma^{+}} \rmd V J^a \LieX A_a^{\mathrm{S}} [ \Sigma^{-} ] - \int_{\Sigma^{-}} \rmd V J^a \LieX A_a^{\mathrm{S}} [ \Sigma^{+} ] \bigg) + \int_\Sigma \xi^b A_b^{\mathrm{S}} J^a \rmd S_a . \label{EDef}
\end{equation}
Bracket notation used here is defined by analogy to \eqref{BracketNot}. The four-dimensional volumes $\Sigma^{-}(s)$ and $\Sigma^{+}(s)$ respectively denote the past and future portions of the body's worldtube with respect to $\Sigma(s)$.

Although this result was most easily motivated by considering
momentum shifts over long times, it is actually valid for any choices
of $s_1$ and $s_2$. It is therefore possible to write
\eqref{FinalDiscrete} in the instantaneous form
\begin{eqnarray}
  \frac{ \rmd }{ \rmd s} (\ItP_\xi + \mathcal{E}_\xi) &=& \int_\Sigma \rmd S_c t^c \bigg[ \frac{1}{2} T^{ab} \LieX g_{ab} + \xi^a J^b  (F_{ab}^{\mathrm{ext}} + F_{ab}^{\mathrm{R}}) \nonumber
  \\
  && \qquad \qquad ~ + \frac{1}{2} \int_W \rmd V' J^a J^{a'} \LieX G_{aa'}^{\mathrm{S}} \bigg]. \label{FinalInst}
\end{eqnarray}
$t^a$ represents the time evolution vector field for the foliation
$\{ \Sigma \}$. This is a general law of motion for compact bodies interacting with electromagnetic fields. Similar methods have been used to obtain a nearly identical result in scalar field theory \cite{HarteScalar}.

\subsection{Self-momentum}
\label{Sect:SelfEn}

The manner in which $\mathcal{E}_\xi$ appears in \eqref{FinalInst} suggests that it may interpreted as an effective momentum associated with the S-type self-field. Another (more obvious) possibility can be obtained from an expression like the one used to define $\ItP_\xi$. Replacing $T^{ab}$ in \eqref{PDef} by the electromagnetic stress-energy tensor \eqref{TEM} associated only with $F_{ab}^{\mathrm{S}}$, let
\begin{equation}
  \mathcal{U}_\xi = \frac{1}{4 \pi} \int_\Sigma g^{cd} (F^{\mathrm{S}}_{ac} F^{\mathrm{S}}_{bd} - \frac{1}{4} g_{ab} g^{fh} F^{\mathrm{S}}_{cf} F^{\mathrm{S}}_{dh}) \xi^a \rmd S^b . \label{CDef}
\end{equation}
Unlike $\mathcal{E}_\xi$, this quantity depends on the details of the field far outside of the body's worldtube. It therefore requires knowledge of the charge distribution in the distant past and future. It also isn't clear that the integral converges if $\Sigma$ is extended to spacelike infinity (even discounting the difficulty in defining $\xi^a$ at large radii).

$\mathcal{U}_\xi$ is therefore unsuitable for use as the effective self-momentum of a dynamical charge distribution\footnote{It is still tempting to try, however. Consider an analog of \eqref{FinalInst} with $\mathcal{E}_\xi$ replaced by $\mathcal{U}_\xi$. For definiteness, let each $\Sigma$ be bounded, so all integrals are manifestly convergent. The force due to $\Lie_\xi G_{aa'}^{\mathrm{S}}$ then disappears. It is replaced by a flux integral through $\partial \Sigma$ and an additional body force requiring knowledge of $\Lie_\xi g_{ab}$ outside of $W$. Both of these terms are difficult to control.}. Despite this, it is known to give reasonable results at least in the stationary case. For simplicity, consider a system in flat spacetime where all variables remain invariant with respect to a timelike Killing field $t^a = \partial / \partial t$. Also suppose that each $\Sigma$ is a hypersurface of constant $t$. It may then be shown that
\begin{equation}
  \mathcal{U}_\xi \rightarrow \int_\Sigma ( \xi^a A^{\mathrm{S}}_a J^b - \frac{1}{2} J^a A_a^{\mathrm{S}} \xi^b ) \rmd S_b \label{EStationary}
\end{equation}
if the potential is in Lorenz gauge. Integrating \eqref{EDef} by parts recovers this expression plus
\begin{equation}
    \frac{1}{2} \int_{\Sigma^{+}} \rmd V \int_{\Sigma^{-}} \rmd V' ( J^{a} \Lie_\xi J^{a'} - J^{a'} \Lie_\xi J^{a} ) G_{aa'}^{\mathrm{S}} \rightarrow 0 .
\end{equation}
$\mathcal{E}_\xi = \mathcal{U}_\xi$ under the given assumptions. The proposed linear and angular momenta of a body's stationary electromagnetic field are exactly as expected.

Returning to the general non-stationary case, this correspondence together with \eqref{FinalInst} provides strong support for the definition of an effective momentum
\begin{equation}
  \hat{\ItP}_\xi = \ItP_\xi + \mathcal{E}_\xi   \label{PEff}
\end{equation}
from which to compute a body's motion under the application of external fields. Deriving (say) a center-of-mass worldline from $\hat{\ItP}_\xi$ rather than $\ItP_\xi$ is closely related to the use of effective masses in more standard treatments of radiation reaction (see, e.g., \cite{OriExtend, HarteEM, Jackson}). Fully taking into account all of the self-field's momentum -- rather than only its inertia -- is essential for understanding the dynamics of general charge distributions.

One reason for this is that the field's energy distribution may have a different center than the matter's. The orbital component of angular momentum effectively forms the first moment of the self-energy distribution. It is especially clear in \eqref{EStationary} that it measures the displacement of the ``center-of-charge'' away from the origin where the associated boost-type Killing fields vanish. In general, mass centers computed from $\ItP_\xi$ and $\hat{\ItP}_\xi$ are therefore not the same. Simple behavior can only be expected to follow from carefully-chosen definitions.

Furthermore, typical mass renormalization techniques that have appeared in the self-force literature effectively assume that the linear momentum associated with the self-field is parallel to that of the matter distribution. There are many interesting systems where this is not the case. As an example, a simple calculation using \eqref{EStationary} shows that the 3-momentum associated with $F_{ab}^{\mathrm{S}}$ can be finite even in the stationary case. Despite the apparent lack of motion, the mechanical momentum needn't vanish either. These two facts are closely related. Stress-energy conservation can be used to show that the mechanical and field momenta are actually negatives of each other in these cases. Choosing any translational Killing field $\xi^a$ transverse to $\Sigma$, $\hat{\ItP}_\xi = 0$. The total 3-momentum of the body as suggested by this formalism therefore has the expected limiting behavior in time-independent systems. Nonzero values for $\ItP_\xi$ by itself have been referred to as ``hidden momenta'' in the literature \cite{Jackson, HiddenMomentum}. Failing to properly take these effects into account when deriving laws of motion could lead to very peculiar predictions. This appears to have been a major reason for the complicated equations of motion derived in \cite{HarteEM}.

Although $\hat{\ItP}_\xi$ avoids these problems, its definition is rather peculiar. Momenta are usually obtained by integrating physical quantities over an appropriate spacelike (or occasionally null) hypersurface. The initial guess \eqref{PDef} for a body's mechanical momentum fits this model. $\mathcal{E}_\xi$ does not. Its definition involves an  additional integral forwards and backwards in time off of the fiducial hypersurface. Computing the momentum at a given time $s$ therefore requires knowledge of the charge distribution in a four-dimensional volume centered on $\gamma(s)$ and with a radius of order the body's proper diameter $D$. If the system of interest does not change significantly on these timescales, it is possible to evaluate the time integrals in \eqref{EDef} explicitly. One is then left with the expected 3-dimensional integrals.

Another brief aside regards the apparent (though fictitious) gauge dependence of $\mathcal{E}_\xi$. It was noted that \eqref{EStationary} holds only in Lorenz gauge. For an arbitrary vector potential generating $F_{ab}^{\mathrm{S}}$, there would be additional terms involving $\nabla^a A_a^{\mathrm{S}}$. Of course, the potentials we have been using are fixed uniquely. There is therefore no gauge freedom at all that our expressions could possibly depend on.

Still, it is interesting to consider ``Detweiler-Whiting type'' Green functions satisfying all of the same properties as the usual ones except for the Lorenz gauge condition. First introduce a two-point tensor field $\chi_a(x,x')$ that vanishes whenever its arguments are timelike-separated. A new Green function $\tilde{G}_{aa'}^{\mathrm{S}}$ can then be obtained from the original via
\begin{equation}
  \tilde{G}_{aa'}^{\mathrm{S}} = G_{aa'}^{\mathrm{S}} + \nabla_{a'} \chi_a + \nabla_a \chi_{a'}.
\end{equation}
This is symmetric, has the same support restrictions as $G_{aa'}^{\mathrm{S}}$, and generates the same field strength $F_{ab}^{\mathrm{S}}$. Attempting to use it in \eqref{EDef} would yield an apparent momentum
\begin{equation}
  \tilde{\mathcal{E}}_\xi = \mathcal{E}_\xi + \frac{1}{2} \int_\Sigma \rmd S_a J^a \int_W \rmd V' J^{a'} \Lie_\xi \chi_{a'} .
\end{equation}
There are no changes if $\nabla_{[a} \Lie_\xi \chi_{b]} = 0$. It is only in this highly restricted sense that our equations can be considered to hold for different Green functions. It would be aesthetically pleasing to do away with the Lorenz condition, although it is not obvious how to do so. Everything is well-defined, so there are no fundamental problems.

\section{Generalized Detweiler-Whiting axiom}
\label{Sect:DetWhit}

It is now possible to derive a generalization of the Detweiler-Whiting (DW) axiom introduced in \cite{DetWhiting}. As remarked above, this postulates that point charges move as though they were test bodies interacting with
\begin{equation}
F^{\mathrm{hom}}_{ab} = F_{ab}^{\mathrm{ext}} + F_{ab}^{\mathrm{R}} \label{Feff}
\end{equation}
via the Lorentz force law. This field is a solution to the homogeneous Maxwell equations. It is almost always smooth at the particle's location, so the DW axiom provides a well-defined prescription for computing trajectories in given external fields. Maxwell electrodynamics is not compatible with the existence of point particles, so this hypothesis cannot be derived in a strict sense.

Physically, however, point particles are usually introduced in order to approximate the motion of sufficiently small extended charge distributions. The DW axiom can therefore be viewed as a statement on the limiting behavior of a certain class of nonsingular objects. Interpreted in this way, it \textit{is} amenable to derivation, and the formalism developed here does so very efficiently. Let the generalized electromagnetic Detweiler-Whiting axiom be the statement that the effective momentum $\hat{\ItP}_\xi$ defined by \eqref{PDef}, \eqref{EDef} and \eqref{PEff} evolves like that of an extended test charge with the same $T_{ab}$ and $J^a$, but moving under the influence of $F_{ab}^{\mathrm{hom}}$ rather than $F_{ab}^{\mathrm{ext}}$. We say that this is true (with respect to $\xi^a$) to order $\epsilon_\xi$ if
\begin{equation}
  \frac{ \rmd }{ \rmd s} \hat{\ItP} _\xi= \int_\Sigma \rmd S_c t^c \left( \frac{1}{2} T^{ab} \LieX g_{ab} + \xi^a J^b  F_{ab}^{\mathrm{hom}} \right) + \epsilon_\xi . \label{FinalDW}
\end{equation}
It is interesting only when $\epsilon_\xi$ can be shown to be negligible compared to other terms here.

Given \eqref{FinalInst}, the critical question therefore becomes the relative importance of
\begin{equation}
  \int_\Sigma \rmd S_c t^c \int_W \rmd V' J^a J^{a'} \LieX G_{aa'}^{\mathrm{S}} . \label{LieTerm}
\end{equation}
Firstly, it is clear that $\LieX G_{aa'}^{\mathrm{S}} = 0$ for any exact Killing field $\xi^a$ that may exist. It follows that $\epsilon_\xi = 0$ in these cases. The generalized DW axiom with respect to any Killing field is exact. This means that there is a sense in which a particular component of a body's momentum is unaffected by $F_{ab}^{\mathrm{S}}$. In maximally symmetric spacetimes -- Minkowski and de Sitter -- all components of linear and angular momentum react only to the regular self-field. This result holds even for charge distributions that may be large and experience rapid internal oscillations. The only approximation used for this conclusion is the assumption that the body does not significantly influence the spacetime geometry\footnote{If it did, exact Killing fields could only exist in stationary and/or axisymmetric systems. Momentum evolution is not particularly interesting in these cases.}.

It is more interesting to ask what happens when no Killing fields exist. $\epsilon_\xi \neq 0$ in these cases, although it can still be negligible. Consider a one-parameter family of matter fields that shrink down to a (fixed) central worldline $\Gamma$ without distorting their shape. Introduce an orthonormal tetrad $e^a_A(x)$ in a neighborhood of these bodies, and let
\begin{eqnarray}
  J^a (\lambda; s,r) = \lambda^{-2} e^a_A (s,r) \tilde{J}^{A} (\lambda, s , r/\lambda) \label{JScale} \\
  T^{ab}(\lambda; s,r) = \lambda^{-2} e^a_{A} (s,r) e^b_{B} (s,r) \tilde{T} ^{A B}(\lambda, s , r/\lambda) \label{TScale}
\end{eqnarray}
for all $\lambda > 0$. $(s,r)$ are Fermi normal coordinates with respect to $\Gamma$. $\tilde{J}^{A}$ and $\tilde{T}^{AB}$ are both assumed to be smooth in all of their arguments. These equations do not change in any essential way under smooth $\lambda$-independent coordinate transformations. Families of this type are essentially the same as those considered in \cite{BobSamMe}. Somewhat more general ones were discussed in \cite{HarteScalar}.

The methods required to estimate \eqref{LieTerm} are very similar to those used for its scalar analog in \cite{HarteScalar}. There, it was shown that at fixed $r/\lambda$ and $r'/\lambda$, Lie derivatives of the various bitensors in \eqref{GExpand} with respect to any GKF $\xi^a$ fall to zero at least as fast as
\begin{equation}
  \LieX \sigma(x,\gamma) |_{\sigma = 0} \sim \Or (\lambda^4), \quad \LieX \ln \Delta(x,\gamma) |_{\sigma = 0} \sim \Or(\lambda^2) .
\end{equation}
near $\Gamma$. Similar methods can easily be used to show that \eqref{Lieg0} also implies that
\begin{equation}
  \LieX g^{a}{}_{a'} |_{\sigma = 0} \sim \Or (\lambda^2) .
\end{equation}
The notation here means that, for example,
\begin{equation}
  \lim_{\lambda \rightarrow 0} \LieX \sigma(x,\gamma)/\lambda^3 = 0 .
\end{equation}
Together, these results imply that \eqref{LieTerm} is $\Or ( \lambda^3 )$. It follows that the error term in \eqref{FinalDW} scales like
\begin{equation}
  \epsilon_\xi \sim \Or(\lambda^3)
\end{equation}
as $\lambda \rightarrow 0$. The Lorentz force due to the external field scales as $\lambda^1$, while the force due to $F_{ab}^{\mathrm{R}}$ scales like $\lambda^2$. For sufficiently small $\lambda$, both of these terms dominate over \eqref{LieTerm}. This shows that the generalized Detweiler-Whiting axiom holds in an interesting sense with respect to all GKFs and all charge distributions described by \eqref{JScale} and \eqref{TScale}.

\section{Motion of small bodies}
\label{Sect:EqsMotion}

It is often useful to simplify an object's motion by considering only the behavior of a single worldline intended to be representative of its overall motion. Several reasonable possibilities for such a mass center have been proposed. Their differences are all expected to be bounded by distances $\delta \gamma \sim (\mathrm{spin \, angular \, momentum}) / (\mathrm{mass})$ \cite{DifferentCMs}. Bodies obeying standard energy conditions likely satisfy $D > \delta \gamma$ \cite{MollerMinRad}, so all potential center-of-mass worldlines should be contained in the convex hull of $W$. There is therefore a sense in which any of them is ``representative'' of the bulk motion. Despite this, a surprisingly large number of potential definitions are excluded simply by demanding that the motion of an uncharged body in flat spacetime be described by a unique geodesic (given standard initial conditions). We adopt a definition -- originally due to Tulczyjew \cite{Tulczyjew} -- that does satisfy this criterion \cite{Dix70a}. It also uniquely specifies both the central worldline $\Gamma$ and the foliation $\Sigma(s)$ in reasonable systems \cite{CM}.

Briefly, we let $\Sigma(s)$ be formed by the intersection of all geodesics that remain orthogonal to $p^a$ -- extracted from $\ItP_\xi$ using \eqref{pToP} -- as they pass through $\gamma(s)$. The boost momentum is also assumed to vanish:
\begin{equation}
  (p^a S_{ab})|_\Gamma = 0. \label{S0iVanish}
\end{equation}
This equation is a direct analog of familiar non-relativistic center-of-mass definitions. In general, the resulting $p^a$ will not be tangent to $\Gamma$. Deviations will usually be very small, although not completely negligible. It is possible to algebraically solve for the center-of-mass velocity $\dot{\gamma}^a$ in terms of $q$, $p^a$, $S_{ab}$, the fields, and the higher multipole moments of $T^{ab}$ and $J^a$. This was accomplished exactly in \cite{EhlRud} in the absence of any electromagnetic fields. A generalization to the present case is not difficult, although the result is rather unwieldy. We will only be using an approximation valid to $\Or(\lambda)$ for the one-parameter family of charge distributions considered in section \ref{Sect:DetWhit}.

It is first convenient to change definitions of the momentum so that all quantities are referred to the map $\hat{\ItP}'_\xi$ introduced in the appendix\footnote{This is not strictly necessary. Multipole expansions for the force and torque associated with $\hat{\ItP}'_\xi$ have already been derived \cite{Dix70a, Dix74, EhlRud}. Being able to use these results simplifies the present derivation. There are also some technical properties of these definitions that are conceptually attractive. Still, exactly the same analysis could be carried out using the original (renormalized) momentum $\hat{\ItP}_\xi$.}. Hats and primes will be omitted for brevity, however. Now  let the worldline parameter $s$ be normalized such that
\begin{equation}
  \dot{\gamma}^a p_a = -m, \label{TimeDef}
\end{equation}
where the mass is defined by letting
\begin{equation}
  p^a = m n^a \label{nDef}
\end{equation}
and $n_a n^a =-1$. These equations imply that $s$ is not precisely a proper time unless $\dot{\gamma}^a = n^a$. With these conventions, the momentum-velocity relation is easily derived using the methods of \cite{EhlRud}. Substituting \eqref{Torque}, it is approximately given by
\begin{equation}
  m \dot{\gamma}_a = p_a + (q S_{a}{}^{c} F^{\mathrm{ext}}_{bc} -2 m Q^{c}{}_{[a} F^{\mathrm{ext}}_{b]c}) n^b/m + \Or(\lambda^3) \label{MomVelFinal}.
\end{equation}
Although one might expect $p^a$ to be exactly $m\dot{\gamma}^a$, it was already remarked in section \ref{Sect:SelfEn} that additional terms must be present. A first approximation for these ``hidden momenta'' appears explicitly here. Such effects are not intrinsically electromagnetic. Similar relations exist even if $J^a=0$ (although they then become nontrivial only at higher orders in $\lambda$).

An immediately useful consequence of \eqref{MomVelFinal} may be derived by contracting it with $\dot{\gamma}^a$. This shows that $\dot{\gamma}^a \dot{\gamma}_a = - 1 + \Or(\lambda^2)$. The worldline parameter $s$ is therefore a proper time up to terms of order $\lambda^2$. These corrections are too small to be relevant to any results derived in this section. The distinction between $s$ and a true proper time may therefore be ignored.

In any case, the full equations of motion can  now be written down very easily. One of these is \eqref{MomVelFinal}. The other two are obtained by combining \eqref{ForceDef}, \eqref{Force} and \eqref{Torque} to yield
\begin{eqnarray}
  \dot{p}_a = (q F^{\mathrm{hom}}_{ab} - \frac{1}{2} S^{cd} R_{abcd}) \dot{\gamma}^b - \frac{1}{2} Q^{bc} \nabla_a F_{bc}^{\mathrm{ext}} + \Or(\lambda^3), \label{pDotFinal}
  \\
  \dot{S}_{ab} = 2 ( p_{[a} \dot{\gamma}_{b]} + Q^{c}{}_{[a} F^{\mathrm{ext}}_{b] c} ) + \Or(\lambda^3). \label{SDotFinal}
\end{eqnarray}
All three equations must be solved simultaneously in order to determine the motion (as well as the mass and angular momentum). They are not complete, however. Appropriate prescriptions must also be given for $Q_{ab}$ and $F_{ab}^{\mathrm{R}}$. In any case, the only effect of the self-field is to act as a small time-dependent shift in the effective external field. The dynamics here are therefore very similar to those of a test body in the dipole approximation.

A better understanding of this system may be gained by introducing (generalized) Fermi derivatives along $\Gamma$ \cite{Dix70a}. For any vector field $k^a$, let
\begin{equation}
  \frac{\mathrm{D}_F k^a}{\rmd s} = \dot{k}^a + 2 \dot{n}^{[a} n^{b]} k_b .
\end{equation}
The Leibniz rule together with $\mathrm{D}_F g_{ab}/ \rmd s = 0$ may used be to generalize this definition for tensor fields of arbitrary rank. Intuitively, it measures changes in components measured with respect to a non-rotating zero-momentum frame.

The Fermi derivative of the angular momentum takes on a very simple form. Before presenting it, note that the center-of-mass condition allows $S_{ab}$ to be replaced by a spin vector
\begin{equation}
  S_a = - \frac{1}{2} \epsilon_{abcd} n^b S^{cd}.
\end{equation}
Inverting, $S_{ab} = \epsilon_{abcd} n^c S^d$. A simple substitution now shows that
\begin{equation}
  \frac{\mathrm{D}_F S_a}{\rmd s} = - \epsilon_{ab}{}^{cd} n^b Q^{f}{}_{[c} F^{\mathrm{ext}}_{d]f} + \Or (\lambda^3) . \label{SpinEvolve}
\end{equation}
The right-hand side represents the ordinary torque expected to act on a test body with electric and magnetic dipole moments. Spin changes absorbed into the definition of the derivative operator are due to Thomas precession. They are necessary in order to preserve $p^a S_a=0$ along $\Gamma$.

Another interesting consequence of the equations of motion is an evolution equation for the mass. $m$ is not usually conserved, but rather varies according to
\begin{equation}
  \frac{\rmd}{\rmd s} \left( m - \frac{1}{2} Q^{ab} F_{ab}^{\mathrm{ext}} \right) = - \frac{1}{2} \frac{\mathrm{D}_F Q^{ab}}{\rmd s} F_{ab}^{\mathrm{ext}} + \Or(\lambda^3).
\end{equation}
One sees that a charge's electromagnetic potential energy directly contributes to its inertia. The right-hand of this equation is related to the internal mechanical work performed by an external field. Phenomenologically, one might consider a dipole moment whose components rigidly rotate with an angular velocity tied to $S_a$ via some (similarly rotating) inertia tensor. It may then be shown that $m$ is given by the sum of a constant ``internal energy,'' the dipole potential energy, and the body's rotational kinetic energy \cite{Dix70a}.

Perhaps the most interesting quantity to calculate is the center-of-mass worldline $\Gamma$. It is difficult to get an intuitive feel for solutions to the coupled system we have derived. Consider instead the derivative of \eqref{MomVelFinal} with appropriate substitutions from \eqref{pDotFinal} and \eqref{SDotFinal}. $n^a$ may be eliminated by replacing it with $\dot{\gamma}^a + \Or(\lambda)$. Making this and related substitutions results in
\begin{eqnarray}
  \frac{ \mathrm{D} }{\rmd s} ( m \dot{\gamma}_a ) = (q F_{ab}^{\mathrm{hom}} - \frac{1}{2} R_{abcd} S^{cd} ) \dot{\gamma}^b - \frac{1}{2} Q^{bc} \nabla_a F_{bc}^{\mathrm{ext}} - S_{ab} \dddot{\gamma}^b \nonumber
  \\
  ~ -2 \left[ Q^{c}{}_{[a} F^{\mathrm{ext}}_{b]c} \ddot{\gamma}^b + \frac{\mathrm{D}_F}{ \rmd s} ( Q^{c}{}_{[a} F^{\mathrm{ext}}_{b]c} \dot{\gamma}^b ) \right] + \Or(\lambda^3) . \label{maFinal}
\end{eqnarray}
The first three terms here are well-known. They are respectively the Lorentz (with self-field), Papapetrou and electromagnetic dipole forces. It is unfortunately rare -- though not unheard of \cite{DeGroot, PenfieldHaus, Jackson} -- to emphasize that such forces represent rates-of-change of momentum rather than direct influences on $m \ddot{\gamma}^a$. Taking into account the aforementioned hidden momenta yields a number of additional terms. These are all test body effects that have nothing to do with self-interaction. All such terms were implicit in Dixon's work \cite{Dix70a}, although they were not displayed in this way. Various other approaches have found similar effects \cite{DeGroot, Jackson, Nyborg, Anandan, GeoPerts}, although there is a surprising amount of disagreement in the literature on relativistic dipole forces. Very recently, however, \eqref{maFinal} was also derived (with radiation reaction) in flat spacetime using a completely independent method \cite{BobSamMe}. Some analysis of its consequences may be found there.

It should be emphasized, however, that \eqref{maFinal} should not be used to indiscriminately replace \eqref{MomVelFinal} and \eqref{pDotFinal}. There is no guarantee that its solutions are physically meaningful over long times. They may not even exist. The implicit assumption that (say) $\ddot{\gamma}^a$ has a well-defined limit as $\lambda \rightarrow 0$ could also break down. At least if $F_{ab}^{\mathrm{R}}$ is ignored, no such problems arise with the original system. The initial value problem has fundamentally changed through the use of substitutions such as $q F_{ab}^{\mathrm{ext}} n^b \rightarrow m \ddot{\gamma}_a + \Or(\lambda^2)$. Similar problems do arise in the original equations of motion if the usual form for the regular self-field is used. This introduces another $\dddot{\gamma}^a$ term multiplied by a small parameter; cf. \eqref{RField} below. One can see that many of the well-known problems of the traditional Abraham-Lorentz-Dirac equation actually arise even in a slightly careless discussion of test particles with significant dipole moments. A simple recipe for avoiding these problems is to use an order reduction procedure. Justifications for this point of view as well as more rigorous applications of perturbation theory may be found in \cite{BobSamMe} and \cite{GeoPerts}.

Putting aside technical issues, we have not yet discussed the self-force in any detail. As can be seen in both \eqref{pDotFinal} and \eqref{maFinal}, this is given by a simple Lorentz-type force associated with $F_{ab}^{\mathrm{R}} (\gamma(s))$. The regular component of the self-field is known to have a well-defined limit even for a point source with finite charge.  Explicitly, it is given by \cite{PoissonRev}
\begin{equation}
  F_{ab}^{\mathrm{R}}  = \frac{2}{3} q \dot{\gamma}_{[a} g_{b]c} ( 2 \dddot{\gamma}^c + R^{c}{}_{d} \dot{\gamma}^d ) + 2 q \lim_{\tau \rightarrow s^{-}} \int_{-\infty}^{\tau}  \nabla_{[a} G^{\mathrm{ret}}_{b]b'} \dot{\gamma}^{b'} \rmd s' . \label{RField}
\end{equation}
The existence of such a limit suggests that the regular self-field of a reasonable extended charge distribution varies negligibly on each time slice $\Sigma(s) \cap W$. We therefore conjecture that \eqref{RField} describes the field of an extended charge distribution up to $\Or(\lambda^2)$. Substituting it into \eqref{maFinal} while reducing order and coupling \eqref{SDotFinal} describes the motion of small self-interacting charge distributions in fixed background spacetimes.

\section{Conclusions}

We have developed a general formalism that greatly simplifies the self-force and self-torque problems for compact charge distributions coupled to Maxwell fields. The central insights regard the effects of the (Detweiler-Whiting) ``singular'' component of a charge's self-field on its motion. As shown by \eqref{FinalInst}, there are two basic mechanisms. First, $F_{ab}^{\mathrm{S}}$ contributes an effective inertia to its source. This component of the field makes the body appear as though it had additional components of linear and angular momentum. These depend only on a quasilocal knowledge of the charge distribution, and cannot generally be mimicked by simple mass shifts.

This is the only direct effect of the singular self-field in maximally symmetric spacetimes. More generally, though, it also exerts a force depending on Lie derivatives of the Green function $G_{aa'}^{\mathrm{S}}$ with respect to the generalized Killing fields. This has a simple interpretation in terms of the degree to which Newton's third law is violated by forces associated with $F_{ab}^{\mathrm{S}}$. All of these conclusions are exact (up to gravitational backreaction effects), and were derived without any reference to specific solutions of Maxwell's equations. They apply for charge distributions of almost arbitrary generality. Only very weak restrictions need to be assumed regarding size, shape, self-energy and internal composition.

Specializing to charge distributions that \textit{are} small and have slow internal dynamics allowed the explicit (though approximate) equations of motion \eqref{MomVelFinal}, \eqref{pDotFinal} and \eqref{SpinEvolve} to be derived. As expected, these involve the DeWitt-Brehme self-force. Electromagnetic and gravitational dipole forces are of similar magnitude in the approximations used, so all effects associated with dipole test bodies are also recovered.

The dynamics are ``universal'' in the sense that they depend on the internal structure only through a handful of simple parameters. A similar statement is unlikely to hold if the equations of motion are expanded to higher orders in $\lambda$. Instead, coefficients would probably arise that cannot be adequately approximated by any finite number of multipole moments. The formalism developed here likely provides a useful starting point for studying such effects. Still, it is not clear how useful such a derivation would be. We have already noted that a number of reasonable definitions for a body's momentum and (and therefore) mass center may proposed. These could all satisfy Detweiler-Whiting axioms in the sense of section \ref{Sect:DetWhit}. It is likely, however, that higher-order self-force and self-torque effects would depend on such choices. It is therefore unclear in what sense they would be physically relevant.

Additional work is also required in order to understand gravitational self-forces in the context of the formalism presented here. This problem is important in its own right, although ignoring it has also constituted a significant gap in the equations of motion presented in section \ref{Sect:EqsMotion}. Dimensionally, gravitational and electromagnetic self-forces should enter at the same order of approximation for families of matter distributions described by \eqref{JScale} and \eqref{TScale}. There could then be nontrivial couplings between the two types of self-interaction. As an example, recent work on the motion of a charged black hole suggests that the force coupling to the background Ricci tensor in \eqref{RField} is modified when gravitational backreaction is taken into account \cite{ChargedBH}. It would be interesting to obtain a more local understanding of this effect.

\appendix

\setcounter{section}{1}

\section*{Appendix: Alternative definitions for the bare momentum}

There is considerable ambiguity in defining the momentum of a charged particle. Many possibilities look at least superficially reasonable while still satisfying a generalized Detweiler-Whiting axiom like the one discussed in section \ref{Sect:DetWhit}. In principle, each element of \eqref{PDef} may be criticized. Temporarily leaving aside the overall form of that equation, definitions for both the body's stress-energy tensor $T^{ab}$ and the generalized Killing fields $\xi^a$ can easily be argued. The only properties of GKFs used here were that they satisfy \eqref{Lieg0} and form a ten-dimensional group which includes any exact Killing fields that may exist. These conditions may be satisfied by vector fields different from those considered in \cite{HarteSyms}, although any alternatives would remain similar on sufficiently small scales.

We have defined the charge's stress-energy tensor by \eqref{TDef}. Again, other reasonable possibilities exist. There is a sense in which only the full stress-energy tensor is physically relevant. Some interesting perturbative statements can be made using this total \cite{BobSamMe}, although  more generally, some split between material and electromagnetic components is essential. The only obvious constraint on the body's stress-energy tensor is that it must vanish outside of $W$. While the electromagnetic stress-energy tensor is unambiguous in vacuum, it could reasonably be given a different form inside matter. It is unclear how the various possibilities would affect the forms of the equations derived in this paper.

Separate from these issues is the basic structure of \eqref{PDef}. It is already strongly suggested by the introduction of an effective momentum $\hat{\ItP}_\xi$ in section \ref{Sect:SelfEn} that $\ItP_\xi$ is incomplete in some sense. A detailed study of the possible multipole moments that can be associated with $T^{ab}$ suggests that $\ItP_\xi$ should be replaced with \cite{Dix70a,Dix74}
\begin{equation}
  \fl \qquad \ItP'_\xi(s) = \ItP_\xi(s) + \int_{\Sigma(s)}  \left( \int_{0}^{1} \kappa^{-1} \nabla_{a'} \sigma (x',\gamma) \xi_{b'}(x') F^{a'b'}(x') \rmd \kappa \right) J^{a}(x) \rmd S_{a}  , \label{PDefDix1}
\end{equation}
It is implicit here that $\gamma = \gamma(s)$. $x' = x'(\kappa)$ is an affinely-parameterized geodesic satisfying $x'(0) = \gamma$ and $x'(1) = x$. In terms of a potential $A_a$, $\ItP_\xi'$ takes on the more intuitive form
\begin{equation}
  \ItP'_\xi = \ItP_\xi + \int_\Sigma \left[ (\xi^a A_a)|^{x}_{\gamma} - \int_{0}^{1} \kappa^{-1} \nabla^{a'} \sigma \LieX A_{a'} \rmd \kappa \right] J^a \rmd S_a . \label{PNEw}
\end{equation}
There is strong evidence that these definitions also arise naturally from Lagrangian considerations \cite{BaileyIsrael}. Note that the $(\xi^a A_a)|_x$ term also appears in the effective momentum $\mathcal{E}_\xi$ defined by \eqref{EDef}.

Although the relation between $\ItP'_\xi$ and the continuum variables $T^{ab}$, $J^a$, $g_{ab}$, and $A_a$ is quite complicated, it has a number of useful properties. There is a simple relation to conservation laws, for example. Temporarily suppose that there exists a vector field $\psi^a$ satisfying $\mathcal{L}_\psi g_{ab} = \mathcal{L}_\psi F_{ab} = 0$. It is then possible to choose a potential $A'_a$ such that $\mathcal{L}_\psi A'_a = 0$. It immediately follows from \eqref{TotForce} that
\begin{equation}
    \frac{\rmd}{\rmd s} [ \ItP_\xi' + q ( \psi^a A'_a )|_{\gamma(s)} ] = 0,
\end{equation}
where $q$ is the body's total charge. The quantity in brackets is conserved for all extended charge distributions. Different choices for $A_a'$ can shift it only by an overall constant \cite{Dix70a}. A special case of this result applicable for pointlike test particles is commonly derived in textbooks \cite{Jackson}.

Another advantage of $\ItP_\xi'$ is that its changes have a particularly simple form in terms of the multipole moments of $T^{ab}$ and $J^a$ \cite{Dix74}. Unfortunately, equivalent expressions are quite complicated when expressed directly in terms of these fields. There is no fundamental obstacle to using \eqref{PDefDix1} instead of \eqref{PDef} in the main development of this paper.  A generalized Detweiler-Whiting axiom like the one derived in section \ref{Sect:DetWhit} will trivially exist for an effective momentum
\begin{equation}
  \hat{\ItP}'_\xi = \ItP'_\xi + \mathcal{E}'_\xi , \label{PHatNew}
\end{equation}
where
\begin{equation}
  \mathcal{E}'_\xi = \mathcal{E}_\xi  - \int_\Sigma \left[ (\xi^a A^{\mathrm{S}}_a)|^{x}_{\gamma} - \int_{0}^{1} \kappa^{-1} \nabla^{a'} \sigma \LieX A^{\mathrm{S}}_{a'} \rmd \kappa \right] J^a \rmd S_a. \label{ENew}
\end{equation}
These definitions are easily applied to a family charge distributions defined by \eqref{JScale} and \eqref{TScale}. Using the results of section \ref{Sect:SelfInt} together with multipole expansions for the force and torque derived in \cite{Dix70a,Dix74}, it may shown that
\begin{eqnarray}
    \hat{F}'_a = q F_{ab}^{\mathrm{hom}} \dot{\gamma}^b - \frac{1}{2} Q^{bc} \nabla_a F_{bc}^{\mathrm{ext}} + \Or(\lambda^3)  , \label{Force}
    \\
    \hat{N}'_{ab} = 2 Q^{c}{}_{[a} F_{b]c}^{\mathrm{ext}} + \Or(\lambda^3) . \label{Torque}
\end{eqnarray}
These quantities are defined by analogy to \eqref{ForceDef}, where the momenta $\hat{p}'_a$ and $\hat{S}'_{ab}$ are associated with $\hat{\ItP}'_\xi$ via a relation like \eqref{pToP}. The homogeneous field $F_{ab}^{\mathrm{hom}}$ has been defined by \eqref{Feff}. Note that terms involving the quadrupole and higher moments of $J^a$ and $T^{ab}$ are negligible in the approximations used here. It may also be shown that the 4-velocity $\dot{\gamma}^a$ never appears anywhere except in the Lorentz force term already written down \cite{Dix74, EhlRud}. This is essential to the use of a center-of-mass condition as part of a well-posed initial value problem.

For reference, the dipole moment of the current distribution $Q_{ab} = Q_{[ab]}(s)$ is given by \cite{Dix70b}
\begin{equation}
  Q_{ab} = \int_\Sigma \nabla_{[a} \sigma \left( \theta_{b]c} \dot{\gamma}^c J^{b'} - \nabla_{b]} \nabla_{a'} \sigma J^{a'} t^{b'} \right) \rmd S_{b'} ,
\end{equation}
where
\begin{equation}
  \theta_{ab}(x,\gamma) = \theta_{(ab)} = \int_{0}^{1} \nabla_a \nabla_{a'} \sigma \nabla_b \nabla^{a'} \sigma \, \rmd \kappa .
\end{equation}
As in \eqref{PDefDix1}, this integral is to be carried out along the geodesic connecting $\gamma$ and $x$. Note that $\theta_{ab} \rightarrow g_{ab}$ in flat spacetime. This remains approximately true for all spacetimes in regions regions where $x$ is sufficiently close to $\gamma$. Given some preferred timelike vector field, $Q_{ab}$ may be decomposed into electric and magnetic dipole moments in the same way that $F_{ab}$ can be written in terms of electric and magnetic fields.

\ack

I am grateful for many helpful comments and discussions with Robert
Wald and Samuel Gralla. This work was supported by NSF grant
PHY04-56619 to the University of Chicago.

\end{document}